\documentclass[prb,twocolumn,superscriptaddress,preprintnumbers,a4paper,amsmath,amssymb,showpacs,floatfix]{revtex4}
\usepackage{graphicx}
\usepackage{bm}

\begin{document}
\title{Magnetic-field induced effects on the electric polarization in \textit{R}MnO$_{3}$ (\textit{R}~=~Dy, Gd)}

\author{R.~Feyerherm}
\email{feyerherm@helmholtz-berlin.de}
\affiliation{Helmholtz-Zentrum Berlin, BESSY, 12489 Berlin, Germany}

\author{E.~Dudzik}
\affiliation{Helmholtz-Zentrum Berlin, BESSY, 12489 Berlin, Germany}

\author{A.~U.~B.~Wolter}
\altaffiliation{now at IFW, 01069 Dresden, Germany}
\affiliation{Helmholtz-Zentrum Berlin, BESSY, 12489 Berlin, Germany}
\affiliation{Institut f\"{u}r Physik der Kondensierten Materie, Technical University Braunschweig, 38106 Braunschweig, Germany}

\author{S.~Valencia}
\affiliation{Helmholtz-Zentrum Berlin, BESSY, 12489 Berlin, Germany}

\author{O.~Prokhnenko}
\affiliation{Helmholtz-Zentrum Berlin, 14109 Berlin, Germany}

\author{A.~Maljuk}
\affiliation{Helmholtz-Zentrum Berlin, 14109 Berlin, Germany}

\author{S.~Landsgesell}
\affiliation{Helmholtz-Zentrum Berlin, 14109 Berlin, Germany}

\author{N.~Aliouane}
\altaffiliation{now at IFE, 2027 Kjeller, Norway}
\affiliation{Helmholtz-Zentrum Berlin, 14109 Berlin, Germany}

\author{L.~Bouchenoire}
\affiliation{XMaS CRG Beamline, European Synchrotron Radiation Facility, 38043 Grenoble, France}

\author{S.~Brown}
\affiliation{XMaS CRG Beamline, European Synchrotron Radiation Facility, 38043 Grenoble, France}

\author{D. N. Argyriou}
\affiliation{Helmholtz-Zentrum Berlin, 14109 Berlin, Germany}

\date{\today}

\begin{abstract}
X-ray resonant magnetic scattering studies of rare earth magnetic ordering were performed on perovskite manganites \textit{R}MnO$_{3}$ (\textit{R} = Dy, Gd) in an applied magnetic field. The data reveal that the field-induced three-fold polarization enhancement for $\mathbf{H}\| a$ ($H \approx 20$~kOe) observed in DyMnO$_{3}$ below 6.5~K is due to a re-emergence of the Mn-induced Dy spin order with propagation vector $\bm{\tau }^{Dy} = \bm{\tau }^{Mn} = 0.385~\mathbf{b}$*, which accompanies the suppression of the independent Dy magnetic ordering, $\bm{\tau }^{Dy} = 1/2~\mathbf{b}$*. For GdMnO$_{3}$, the Mn-induced ordering of Gd spins is used to track the Mn-ordering propagation vector. The data confirm the incommensurate ordering reported previously, with $\bm{\tau }^{Mn}$ varying from 0.245 to 0.16 $\mathbf{b}$* on cooling from $T_{N}^{Mn}$ down to a transition temperature $T^\prime$. New superstructure reflections which appear below $T^\prime$ suggest a propagation vector $\bm{\tau }^{Mn} = 1/4~\mathbf{b}$* in zero magnetic field, which may coexist with the previously reported A-type ordering of Mn. The Gd spins order with the same propagation vector below 7~K. Within the ordered state of Gd at $T~$=~1.8~K we find a phase boundary for an applied magnetic field $\mathbf{H}\| b$, $H = 10$~kOe, which coincides with the previously reported transition between the ground state paraelectric and the ferroelectric phase of GdMnO$_{3}$. Our results suggest that the magnetic ordering of Gd in magnetic field may stabilize a cycloidal ordering of Mn that, in turn, produces ferroelectricity.
\end{abstract}

\maketitle

\section{Introduction}

There is a great current interest in multiferroic materials, which are characterized by the coexistence of ferroelectric and magnetic ordering.\cite{Spaldin:2005,Cheong:2007} Widely studied key examples for such magneto-electric multiferroics are the distorted perovskites \textit{R}MnO$_{3}$, where \textit{R}~=~Tb, Dy, or Gd. One of the most spectacular effects, which shows the strong magneto-electric coupling in these materials, is the magnetic-field induced flop of the electric polarization direction from $\mathbf{P}\| c$ to $\mathbf{P}\| a$ observed in Tb- and DyMnO$_{3}$. For TbMnO$_{3}$ it has been suggested that this polarization flop may be driven by the interplay between Mn and Tb,\cite{Kimura:2003,Kimura:2005} which has proven to be strong,\cite{Prokhnenko:2007} but whose microscopic origin is still under debate.

Current theoretical models relate the ferroelectricity in the \textit{R}MnO$_{3}$ series to a complex cycloidal magnetic ordering of the Mn spins due to magnetic frustration effects.\cite{Katsura:2005,Mostovoy:2006,Kenzelmann:2005} In such cycloidal structures the inverse Dzyaloshinsky-Moriya interaction, resulting from spin-orbit coupling on the Mn ions, is the driving force of polar lattice distortions.\cite{Xiang:2008,Malashevich:2008} In these models, any contribution from the rare earth \textit{R} is usually neglected. However, several observations suggest that the rare earth have a significant influence on the ferroelectric properties of the \textit{R}MnO$_{3}$. In the series (Eu,Y)MnO$_{3}$, for example, where Eu and Y are non-magnetic, the electric polarization $\mathbf{P}$ is basically oriented along the $a$ axis.\cite{Hemberger:2007,Ivanov:2006} On the other hand, strongly anisotropic magnetic rare earths like Tb and Dy appear to affect the electric polarization state: in Tb- and DyMnO$_{3}$ the orientation of the polarization is $\mathbf{P}\| c $ in zero magnetic field. In the less anisotropic GdMnO$_{3}$, electric polarization $\mathbf{P}\|  a$ exists only in applied magnetic field. Finally, it has been demonstrated that the enhancement\cite{Goto:2004} of the ferroelectric polarization of DyMnO$_{3}$ is caused by a Mn-induced Dy spin order as long as Dy is not ordered independently.\cite{Prokhnenko:2008} These observations point to a mutual influence between the \textit{R} and the Mn that, in turn, determines the ferroelectric properties.\cite{Aliouane:2008}

Despite the importance of the \textit{R} in determining the magnetic and ferroelectric properties in the \textit{R}MnO$_{3}$ series, many details of the \textit{R} magnetic ordering and its interplay with the Mn ordering are still unclear. The magnetic ordering for Dy and especially Gd compounds are difficult to study with neutron scattering techniques because of the high neutron absorption cross-sections for these elements. Synchrotron x-ray diffraction (XRD) and x-ray resonant magnetic scattering (XRMS) have proven to be suitable techniques to fill this information gap.

The \textit{R}MnO$_{3}$ (\textit{R} = Gd, Tb, Dy) undergo various stages of magnetic ordering of the \textit{R} and Mn moments as function of $T$ (see Table 1). At the Neel temperature $T_{N}^{Mn}$ a Mn sinusoidal order arises with a gradual shift of the propagation vector $\tau ^{Mn}$ as $T$ is reduced.\cite{Kimura:2005,Kimura:2004} At a transition temperature $T^\prime$ $\tau ^{Mn}$ stabilizes in Tb- and DyMnO$_{3}$ simultaneously with the emergence of a spontaneous electric polarization $\mathbf{P}\| c$. For GdMnO$_{3}$ a lock-in transition to A-type ordering             ($\bm{\tau }^{Mn}$ = 0) has been reported.\cite{Arima:2005} This compound remains paraelectric in the absence of a magnetic field.

\begin{table*}[tb]
\begin{center}
\caption{Ordering temperatures and propagation vectors of the various magnetic states in the \textit{R}MnO$_{3}$ series, \textit{R}~=~Gd, Tb, and Dy. The data were taken from refs. \cite{Kimura:2005,Kimura:2004} except where 
indicated; $^{\S }$our previous work \cite{Prokhnenko:2007,Prokhnenko:2008,Feyerherm:2006}; $^{\S \S }$this work. All 
propagation vectors $\tau $ are along $\mathbf{b}$*. We assume that $\tau^{Mn} = \tau ^{R}$ below $T_{N}^{R}$ for GdMnO$_{3}$, although there is no direct evidence.}

\begin{tabular}{|c|c|c|c|c|c|c|c|}
\hline
comp.& 
$T_{N}^{Mn}$& 
$\tau ^{Mn}$ \par ($T_{N}^{Mn} > T > T^\prime)$& 
$T^\prime$& 
$\tau ^{Mn}$ \par ($T^\prime > T > T_{N}^{R})$& 
$T_{N}^{R}$& 
$\tau ^{Mn}$ \par ($T < T_{N}^{R})$ & 
$\tau ^{R}$ \\
\hline
GdMnO$_{3}$& 
43 K& 
0.24...0.2& 
23 K& 
0, 1/4 $^{\S \S }$& 
7 K& 
0, 1/4 (?) $^{\S \S }$& 
1/4 $^{\S \S }$ \\
\hline
TbMnO$_{3}$& 
42 K& 
0.285...0.275& 
27 K& 
0.275& 
7 K& 
2/7 $^{\S }$& 
3/7 $^{\S }$ \\
\hline
DyMnO$_{3}$& 
39 K& 
0.36...0.385& 
18 K& 
0.385& 
6.5 K $^{\S }$& 
0.405 $^{\S }$& 
1/2 $^{\S }$ \\
\hline
\end{tabular}
\label{tab1}
\end{center}
\end{table*}

In DyMnO$_{3}$ a commensurate Dy ordering with propagation vector $\bm{\tau }^{Dy}= 1/2~\mathbf{b}$* (with G$_{x}$A$_{y}$-mode stacking in Bertaut's notation\cite{Bertaut:1963}) is observed below 6.5 K. This commensurate ordering is accompanied by an incommensurate lattice modulation with a period $q = \tau ^{Dy} \pm \tau ^{Mn}$ which is induced by a magnetoelastic coupling between Dy and Mn magnetic moments $\propto \bm{\mu}_{Dy}\cdot \bm{\mu}_{Mn}$.\cite{Feyerherm:2006} No information on the Gd propagation vector in GdMnO$_{3}$ has been reported to date. In this work we show that below $T_{N}^{Gd}$ = 7 K the Gd moments order commensurately with $\bm{\tau }^{Gd~} = 1/4~\mathbf{b}$*. (Since all observed magnetic propagation vectors are along $\mathbf{b}$*, in the following we will give $\tau $ values only as numbers).

Both Dy- and GdMnO$_{3}$ exhibit interesting effects induced by external magnetic fields. Below 10 K DyMnO$_{3}$ shows a significant enhancement of the electric polarization $\mathbf{P}\| c$ by a factor of up to 3.5 for magnetic fields between 10 and 50~kOe applied along $a,$ and between 10 and 20~kOe for fields along $b$. In the ground state of GdMnO$_{3}$ ($T << T_{N}^{Gd}$), in turn, an electric polarization  $\mathbf{P}\| a$ is observed only when a magnetic field $H > 10$~kOe is applied along $b$.

In this report we present XRMS results of the \textit{R} magnetic ordering in an applied magnetic field $\mathbf{H}\|  a$ for DyMnO$_{3}$ and $\mathbf{H}\| b$ for GdMnO$_{3}$ which give an additional physical insight on the observed field-induced effects.

\section{Experimental}

High quality single crystals of Dy- and GdMnO$_{3}$ were grown by the crucible-free floating zone method in argon atmosphere. The as-grown crystals were characterized by x-ray Laue diffraction. Magnetization measurements, carried out with a commercial PPMS system, show excellent agreement with published data. XRMS measurements were performed on the MAGS beamline operated by the Helmholtz-Zentrum Berlin at the BESSY synchrotron source and, where indicated, on the XmaS beamline at the ESRF, Grenoble. Both beamlines can be equipped with superconducting magnets which allow magnetic fields to be applied perpendicularly to the scattering plane. At XmaS the maximum available field is $H_{max} = 40$~kOe and the base temperature $T_{min} = 1.8$~K. At MAGS, $H_{max} = 50$~kOe and $T_{min} = 1.6$~K in zero field or $T_{min} = 4$~K with the magnet. The beamline magnet at MAGS is one of the first in the world to be based on high-$T_{c}$ superconductor technology.\cite{Pooke:1}

The magnets set constraints to the possible experimental geometries. In DyMnO$_{3}$ we investigated the behavior for scattering vectors along $\mathbf{b}$* with the field direction $\mathbf{H}\| a$. For GdMnO$_{3}$, in order to apply a magnetic field approximately along the $b$ axis, a crystal was cut and aligned to allow scattering vectors close to $\mathbf{c}$*. With this latter geometry only a single magnetic Bragg reflection could be studied. Measurements were performed using photon energies close to the Dy L$_{3}$ and Gd L$_{2}$ edges at 7.79 and 7.93~keV, respectively. The x-ray beam had a degree of linear polarization of $\approx 85$~\%. Linear polarization analysis of the scattered beam was carried out using the (006) reflection of a HOPG crystal. All XRMS data presented below were measured in $\sigma\rightarrow\pi^{\prime}$ geometry. We have reduced local sample heating by the x-ray beam by using appropriate absorber foils where feasible. However, small offsets of the temperature scale by $\approx 1$~K between different experimental data sets are possible, especially when measuring weak intensities. More details of the general experimental procedure have been reported previously.\cite{Prokhnenko:2008}

\section{Results and discussion}

\subsection{DyMnO$_{3}$}

\begin{figure}[bt!]
\begin{center}
\includegraphics[scale=0.7]{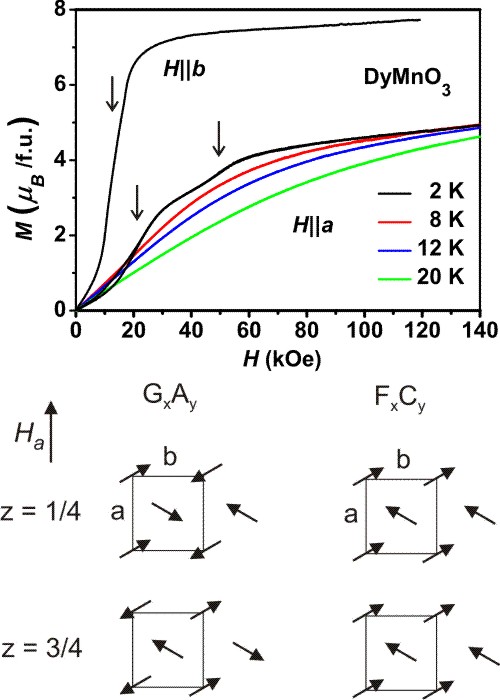}
\caption{(Color online) (top) Magnetic field dependence of the magnetization of DyMnO$_{3}$ measured at various temperatures for applied fields $\mathbf{H}\| a$ and at 2~K for $\mathbf{H}\| b$, arrows mark inflection points associated with metamagnetic transitions; (bottom) moment stacking for the low-field G$_{x}$A$_{y }$ and high-field forced ferromagnetic alignment of the $a$ component (F$_{x}$C$_{y})$ for $\mathbf{H}\| a$; the unit cell contains two Dy layers at $z =$~1/4 and 3/4 with Dy at (0.983, 0.084, 1/4) and equivalent positions.}
\label{fig1}
\end{center}
\end{figure}

Magnetization data for DyMnO$_{3}$, measured at various temperatures for applied magnetic fields $\mathbf{H}\| a$ up to 140~kOe, are shown in Fig.~\ref{fig1}. At $T = 2$~K, i.e., below the Dy ordering temperature, a two-step metamagnetic behavior is observed with a first transition around 20~kOe, and a second transition around 50~kOe (inflection point positions). Above about 60~kOe the $a$-component of the Dy moment appears to be ferromagnetically aligned. The magnetization reaches a value of $\sim 4.5 \mu _{B}$/f.u. at 100~kOe. Magnetization data for $\mathbf{H}\| b$ has a single-step metamagnetic transition around 15~kOe and a magnetization of $\sim 7.5~\mu _{B}$/f.u. at 100~kOe. From these data one can infer Ising anisotropy with an easy direction oriented within the \textit{ab} plane about 30$^\circ$ from the $b$ axis. This behavior is similar to that reported for related compounds like DyAlO$_{3}$.\cite{Cashion:1998} Due to the specific symmetry of the \textit{R}MnO$_{3}$ (space group \textit{Pbnm}), this local easy axis is staggered for a pair of nearest neighbor Dy. We tentatively associate the two steps in the magnetization process for $\mathbf{H}\| a$ with a change of the stacking modes of Dy moments according to the sequence G$_{x}$A$_{y}$ $\to $ intermediate state $\to $ F$_{x}$C$_{y}$, see Fig.~\ref{fig1} (bottom), while for $H\| b$ the high-field state is C$_{x}$F$_{y}$.

\begin{figure}[bt!]
\begin{center}
\includegraphics[scale = 0.7]{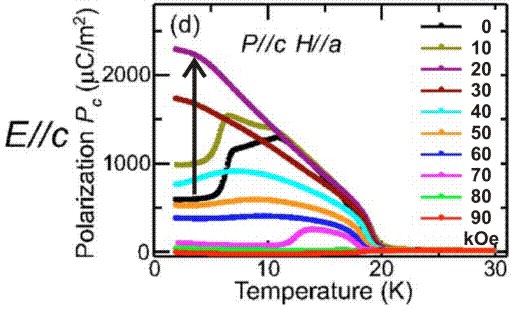}
\caption{(Color online) Temperature dependence of the electric polarization $\mathbf{P}\| c$ measured at various applied magnetic fields $\mu_{0} \mathbf{H}\| a$ (after Ref.~4). A threefold increase of $\mathbf{P}\| c$ occurs when applying a magnetic field of 2~T below 5~K (see arrow).}
\label{fig2}
\end{center}
\end{figure}

It is instructive to compare this magnetization behavior with previously reported electric polarization data (Fig.~\ref{fig2})\cite{Kimura:2005}. For zero magnetic field the electric polarization shows a sudden increase on heating above the AFM transition temperature of Dy, $T_{N}^{Dy}$ = 6.5 K. We have recently reported\cite{Prokhnenko:2008} that this behavior is due to an enhancement of the electric polarization related to the Mn-induced magnetic ordering of Dy moments with propagation vector $\tau ^{Mn}$. From neutron diffraction we have derived a magnitude of the induced Dy moment of 2.5~$\mu_{B}$ at 10~K, principally aligned along the $b$ axis. This induced ordering is only present in the temperature range $T_{N}^{Dy} < T < T^\prime$. Below $T_{N}^{Dy}$ it is replaced by the independent Dy ordering with $\tau ^{Dy} = 1/2$ which does not contribute to the electric polarization. For $\mathbf{H}\| a \ge 20$~kOe the electric polarization data show no kinks. This might suggest that the independent Dy ordering with $\tau^{Dy} = 1/2$ is completely suppressed by a magnetic field $\mathbf{H}\| a \ge 20$~kOe. 

\begin{figure}[bt!]
\begin{center}
\includegraphics[scale = 0.7]{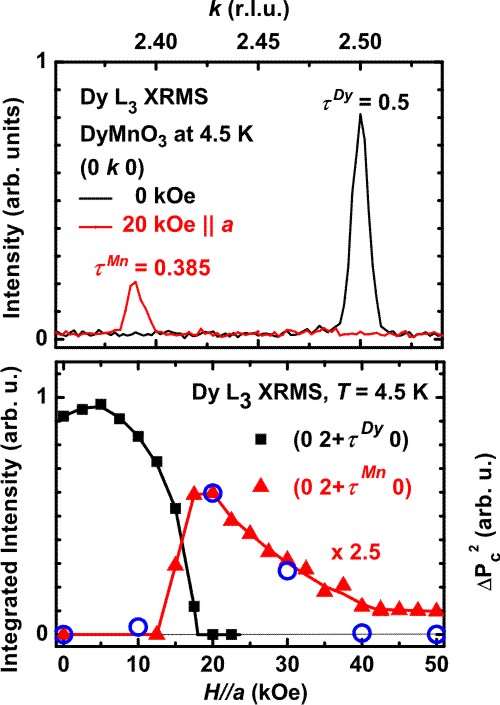}
\caption{(Color online)(top) XRMS $k$-scans along (0~$k$~0) measured at 4.5~K in zero field and in 20~kOe applied magnetic field $\mathbf{H}\| a$; (bottom) Magnetic field dependence of the integrated intensities of the (0~2.5~0) and 
(0~2.385~0) reflections at 4.5~K. Open circles show the square of the polarization enhancement $\Delta P_{c}$ defined within the text. The scaling factor is fixed to give a match at 20~kOe. Lines in the bottom figure are guides to the eye.}
\label{fig3}
\end{center}
\end{figure}

To verify this hypothesis, DyMnO$_{3}$ was studied by XRMS at the Dy-L$_{3}$ resonance in magnetic fields $\mathbf{H}\| a$ with scattering vectors (0~$k$~0). First, $k$-scans were measured at 4.5~K in zero field and in 20~kOe, with $k$ values around $2 + \tau $ (Fig.~\ref{fig3}). The Bragg reflection related to the individual Dy ordering $\tau ^{Dy} = 1/2$ does vanish in an applied field of 20~kOe. Simultaneously, another reflection with $\tau ^{ Dy} = 0.385$ appears. This value coincides with the abovementioned Mn-induced Dy ordering with $\tau ^{Mn}$. We conclude that the field-induced suppression of the $\tau ^{Dy} = 1/2$ ordering is accompanied by a re-emergence of the Mn-induced ordering and is directly linked to the field-induced increase of the electric polarization below 6.5~K at 20 and 30~kOe (see Fig.~\ref{fig2}).

From the magnetic field dependence of the (0~2+$\tau ^{Dy}$~0) reflection intensity at base temperature (Fig.~\ref{fig3}) we obtain a critical field of 18~kOe for the transition of the Dy ordering from $\tau ^{Dy} =  1/2$ to $\tau ^{Mn} = 0.385$. The Bragg reflection related to $\tau ^{Mn} = 0.385$ has its maximum intensity around 20~kOe and decreases rapidly for larger fields. Thus the Mn-induced ordering of Dy is gradually suppressed for magnetic fields above 20~kOe. Above 40~kOe the intensity levels off, probably marking a second transition. It is worth mentioning that the position of the (0~2+$\tau ^{Mn}$~0) reflection does not vary significantly with the strength of the applied magnetic field (not shown), which implies that the Mn propagation vector is field-independent, similar to the behavior reported for $\mathbf{H}\| b$.\cite{Strempfer:2007}

These observations confirm that the first metamagnetic step in the magnetization data (Fig.~\ref{fig1}) is indeed related to a breakdown of the independent Dy ordering. Above this transition, however, the Dy moments still carry an antiferromagnetic (AF) modulation with the same propagation vector as the Mn-induced ordering. The magnetization can only be saturated (forced ferromagnetic alignment) when this modulation is also suppressed. This point is reached above the second transition at $\sim 50$~kOe. It is worth noting that this field value is close to the polarization flop transition from $\mathbf{P}\| c$ to $\mathbf{P}\| a $ observed for $\mathbf{H}\| a > 65$~kOe.\cite{Kimura:2005}

It is also interesting to compare the magnetic field dependence of the induced ordering (Fig.~\ref{fig3}) with the field dependence of the electric polarization (Fig.~\ref{fig2}). Obviously, there is a direct correspondence between the intensity of the induced-ordering related Bragg reflection and the magnitude of the electric polarization at 4.5 K. We performed a quantitative analysis based on the assumption that the $\mu_{y}$ and $\mu_{z}$ components of the Mn-cycloid stay unchanged (or change only slightly) under application of a magnetic field. Then the field-dependent electric polarization enhancement ($\Delta P_{c}(H)= P_{c}(H) - P_{0}$) should be proportional to the size of the induced Dy moment, justified by the model presented in our previous report.\cite{Prokhnenko:2008}. Since the related Bragg intensity is proportional to the square of the induced moment, the square of the polarization enhancement should scale with the (0~2+$\tau ^{Mn}$~0) Bragg intensity. The experimental data (solid circles in Fig.~\ref{fig3}) follow this scaling well except for $H \ge 40$~kOe. One possible explanation for the mismatch in this field-region is that the above assumption becomes invalid when approaching the flop of the electric polarization at $H \sim 65$~kOe, since this is related to changes on the Mn-sublattice.

\begin{figure}[bt!]
\begin{center}
\includegraphics[scale = 0.7]{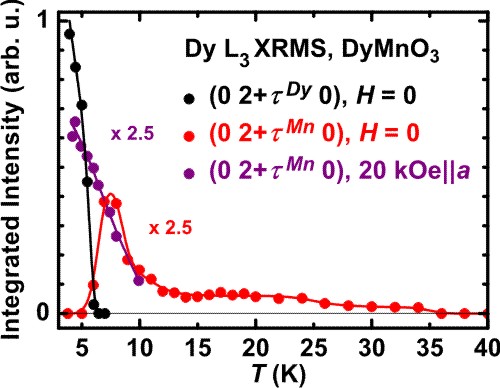}
\caption{(Color online) Temperature dependence of the (0~2+$\tau ^{Dy}$~0) integrated intensity in zero field (black curve) and that of the (0~2+$\tau ^{Mn}$~0) reflection in zero field (red) and 20kOe applied field $\mathbf{H}\| a$ (purple). Lines are guides to the eye.}
\label{fig4}
\end{center}
\end{figure}

To complete the discussion of the XRMS data on DyMnO$_{3}$, Fig.~\ref{fig4} shows the temperature dependence of the (0~2+$\tau ^{Mn}$~0) reflection in zero field and 20~kOe together with the (0~2+$\tau ^{Dy}$~0) reflection in zero field. We derive a zero-field transition temperature of 6.5~K for the Dy ordering, consistent with our previous findings.\cite{Prokhnenko:2008} Interestingly, in an applied field of 20~kOe, the intensity of the (0~2+$\tau ^{Mn}$~0) reflection increases continuously on cooling. At 4.5 K it reaches a value five times larger than at 10 K. We relate this intensity increase with the continuous increase of the electric polarization on cooling in a 20~kOe field (Fig.~\ref{fig2}). These data also follow roughly the expected scaling of the induced Bragg peak intensity with the square of the electric polarization enhancement.

In contrast to our previous report, we observe a residual intensity above $T^\prime = 18$~K of the (0~2+$\tau ^{Mn}$~0) reflection (F-mode, $h+k$ even and $l$ even, in Bertaut's notation) in zero field in the present data (Fig.~\ref{fig4}), which vanishes only above 36~K, i.e., close to $T_{N}^{Mn}$. We ascribe this apparent discrepancy to the different azimuth orientations in the two experiments, namely $a$ perpendicular to the scattering plane in the present, but $a$ parallel to the scattering plane in the previous experiment. Since the XRMS amplitude is a measure of the projection of the ordered moment onto the scattered beam, the present geometry highlights XRMS from any $c$ component of the Dy moment, which was suppressed in the previous experiment. Thus, the present observation of residual (0~2+$\tau ^{Mn}$~0) intensity above 18~K suggests that Dy -- besides the previously reported component of maximally 2.5~$\mu_{B}$ along the $b$ axis\cite{Prokhnenko:2008} -- carries a small induced moment component parallel to the $c$ axis above $T^\prime$ which, however, does not induce an electric polarization in that temperature range. Similar observations have also been made by x-ray resonant scattering in TbMnO$_{3}$.\cite{Mannix:2007,Voigt:2007} 

Such a residual $c$ component may be also present at lowest $T$ in applied fields and thus account for the residual (0~2+$\tau ^{Mn}$~0) intensity observed in the field-dependent data at $H \ge 40$~kOe (Fig.~\ref{fig3}). However, in such a picture it would be difficult to understand why the $b$ component of the induced moment -- and with it the polarization enhancement -- is suppressed by magnetic field $\mathbf{H}\| a$ but the $c$ component is not. For TbMnO$_{3}$ it has been argued that the residual Mn-induced polarization of the Tb in the paraelectric state solely resides on the Tb 5$d$ orbitals and that 4$f$ states are not affected.\cite{Mannix:2007,Voigt:2007} Therefore, such a contribution to the spin polarization may survive even when the 4$f$ ordering is suppressed.

To conclude the discussion of DyMnO$_{3}$, we have shown that for $\mathbf{H}\| a$ the intermediate field region with 
enhanced electric polarization for $T < 6.5$~K is characterized the re-occurrence of an induced Dy ordering with $\tau ^{Mn}$. Only for fields large enough to fully suppress the AFM arrangement of Dy moments with $\tau ^{Mn}$, the enhancement of the electric polarization by the Dy vanishes. All observations of the present magnetic-field dependent study confirm our previous conjecture\cite{Prokhnenko:2008} that the Mn-induced ordering of Dy is responsible for any enhancement of the electric polarization in DyMnO$_{3}$.

\subsection{GdMnO$_{3}$}

\begin{figure}[bt!]
\begin{center}
\includegraphics[scale = 0.7]{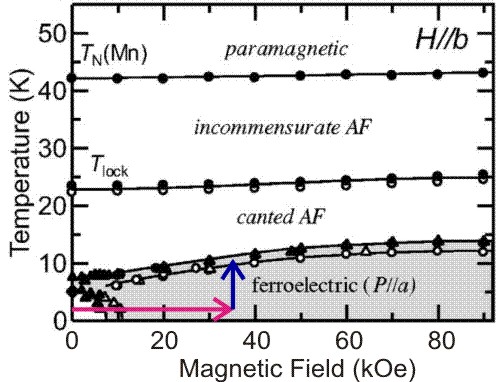}
\caption{(Color online) Magneto-electric phase diagram of GdMnO$_{3}$ with magnetic field along the $b$ axis determined from various experimental techniques (after Ref.~4). The gray region indicates the ferroelectric phase. The arrows indicate the paths of the temperature and field scans shown in Fig.~\ref{fig8}.}
\label{fig5}
\end{center}
\end{figure}

In contrast to Tb- and DyMnO$_{3}$, only very limited data on the magnetic ordering in GdMnO$_{3}$ have been reported so far. The only available information comes from XRD experiments, in which second-order Bragg reflections from 2$\tau ^{Mn}$ lattice modulations induced by magnetic ordering were studied.\cite{Kimura:2004,Arima:2006} From these data it was deduced that the Mn magnetic moments ordered incommensurately below $T_{N}^{Mn} =43$~K. Below the lock-in temperature, $T^\prime = 23$~K, the associated Bragg reflections vanished. This was interpreted as a transition to A-type antiferromagnetic (AF) ordering of the Mn ($\tau ^{Mn} = 0$), associated with weak ferromagnetism (canted AF). The application of a magnetic field $H > 10$~kOe along the $b$ axis at $T = 8$~K lead to the appearance of new superstructure Bragg reflections related to a propagation vector $\tau = 1/4$. This was explained as a field-induced change of the Mn magnetic ordering from the canted AF state to a moment stacking with $\tau ^{Mn} = 1/4$,\cite{Arima:2006} associated with the transition to the ferroelectric phase. The corresponding phase diagram is shown in Fig.~\ref{fig5}. 
 
The results discussed in the following, however, shed new light on the Mn magnetic ordering and the magnetic origin of the ferroelectric phase. Firstly, we present data taken in zero magnetic field to determine the Gd magnetic ordering in the ground state and its temperature dependence. We will show that -- in a certain temperature range -- the Mn-induced ordering of the Gd moments can be used to track the Mn propagation vector. Secondly, we discuss in-field data that provide evidence that the phase boundaries of the ferroelectric phase and field-dependent magnetic phases of Gd coincide.

\subsubsection{Zero-field measurements}

Magnetic ordering of Gd has been reported to occur below 6.5~K.\cite{Hemberger:2004} Nevertheless, no information on the Gd ordering propagation vector have been reported yet. To study the ground-state Gd magnetic ordering, we carried out a reciprocal space survey in zero field at $T =1.6$~K, using XRMS close to the Gd L$_{2}$ edge. It reveals the presence only of A- and F-mode superstructure reflections (0~$k\pm \tau ^{Gd}$~$l$) with $\tau ^{Gd} = 1/4$, ($k$ even and $l$ odd or even, respectively). Polarization analysis shows that in $\sigma\rightarrow\sigma^{\prime}$ geometry these reflections are suppressed, an indication of their magnetic character. Variation of the x-ray energy through the Gd L$_{2}$ absorption edge proves their resonant character, with a strong resonant enhancement of the Bragg intensity by a factor $\approx $ 100 at the edge. We therefore associate these Bragg reflections with a magnetic ordering of the Gd moments with propagation vector $\tau ^{Gd} = 1/4$.

\begin{figure}[bt!]
\begin{center}
\includegraphics[scale = 0.70]{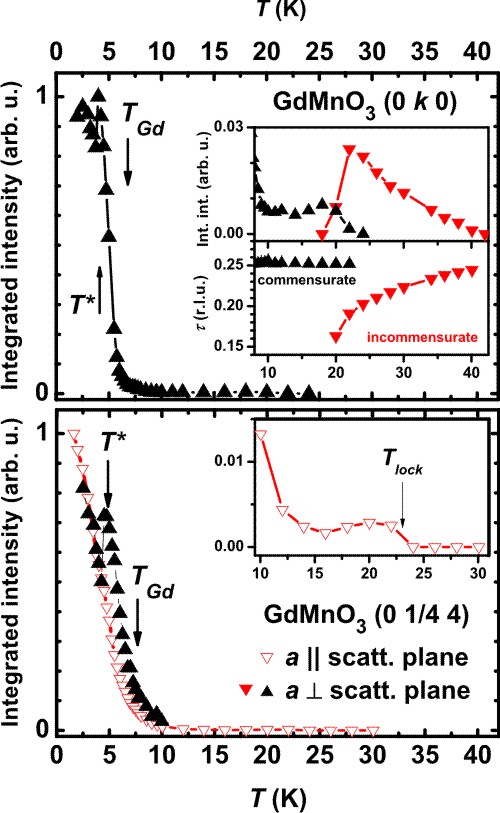}
\caption{(Color online) Temperature dependence in zero field of the integrated intensities of the (0~3.75~0) (top) and the (0~0.25~4) reflections (bottom). The inset in the top shows a blow-up of the $T$ region above 8~K. Incommensurate 
reflections of type (0~$k$~0) are observed above $T^\prime$. The temperature dependence of $k$ is shown in the lower portion of the inset. The bottom figure shows the integrated intensity the (0~0.25~4) reflection for two different azimuth orientations, namely a parallel (open symbols) and a perpendicular to the scattering plane (solid symbols). The data for the different azimuth orientations are normalized to match below $T$*. The inset shows a blow-up of the $T$ region above 8~K. Here, incommensurate reflections were not measured. All data were taken on heating. A slight offset between the temperature scales of the top and bottom figures is probably caused by excessive heating of the sample by the x-ray beam in the former experiment.}
\label{fig6}
\end{center}
\end{figure}

In order to determine the orientation of the ordered Gd moment, we measured the azimuth dependence of the (0~3.75~0) reflection in zero field at 1.6~K. On rotation of the scattering plane from \textit{bc} to \textit{ab}, the integrated intensity of this reflection, normalized to the (0~4~0) reflection, is reduced down to $\approx 10$~\% (not shown). This indicates that in zero field the Gd ordered moment \textit{$\mu$} is principally aligned along the $c$ direction, consistent with magnetization measurements which show that $c$ is the easy magnetic axis for Gd.\cite{Hemberger:2004} The residual intensity observed for the scattering plane \textit{ab}, however, is significantly larger than expected from the limited degree of polarization, since with the (0~4~0) reflection a $\pi\rightarrow\pi^{\prime}$ cross-talk in the $\sigma\rightarrow\pi^{\prime}$ setting of only 3~{\%} was determined. This observation points to an additional component of the Gd moment perpendicular to $c$ at 1.6~K. Taking into account the projection of the ordered moment onto the scattered beam ($\theta \approx 30^\circ$), the residual intensity is consistent with  $\mu_{a} \approx 0.25~\mu_{c}$ or $\mu_{b}$ $\approx 0.45~\mu_{c}$.

\begin{figure}[bt!]
\begin{center}
\includegraphics[scale = 0.7]{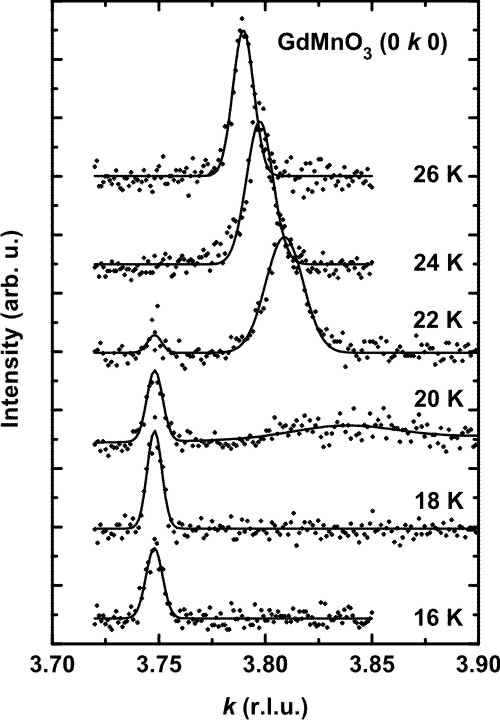}
\caption{(Color online) Selection of reciprocal space scans along (0~$k$~0) around $k = 3.75$ for various temperatures close to $T^\prime$. The solid lines are fits to the data from which the parameters plotted in Fig.~\ref{fig6} (top, inset) were derived. In a crossover region commensurate ($\tau =1/4$) and incommensurate reflections coexist, consistent with a first-order transition at $T^\prime$. In this experiment the value of $T^\prime$ is slightly lower than in the previous report, probably because of heating of the sample by the x-ray beam.}
\label{fig7}
\end{center}
\end{figure}

The temperature dependence of the (0~3.75~0) intensity was measured with the scattering plane parallel to \textit{bc} (Fig.~\ref{fig6}, top). Interestingly, after a small decrease, its intensity shows a sudden increase by about 20\% upon heating through $T$* = 4.5~K. On further heating, the intensity drops rapidly, marking the Gd ordering transition around $T_{N}^{Gd} = 7$~K. Similar temperature dependences have also been observed for other Bragg reflections, like (0~1.75~3) and (0~2.25~4) (not shown) and (0~0.25~4) (Fig.~\ref{fig6}, bottom). It is interesting to note that such a sequence of two transitions has been reported before by Kimura \textit{et al.}\cite{Kimura:2005} (Fig.~\ref{fig5}). In their report the \textit{lower} transition (at 5.1~K) was assigned to Gd ordering. Our results, in contrast, suggest that the \textit{upper} transition (at 8.2~K) is the one to be associated with Gd ordering (with slightly lower values for the transition temperatures in our sample). The lower transition is presumably related to a modification of the magnetically ordered structure. It is worth noting that in the $T$ range $T$*$ < T < T_{N}^{Gd}$ GdMnO$_{3}$ was found to be weakly ferroelectric ($\mathbf{P}\| a)$ in zero magnetic field.\cite{Kimura:2005}

Interestingly, the (0~3.75~0) reflection does not vanish completely above $T_{N}^{Gd}$, but can be tracked up to about  $T^\prime$ with an intensity of less than 1\% of the 2~K value in the temperature range $10~{\rm K} < T < T^\prime$.  We checked that this residual intensity vanishes when the x-ray energy is tuned 20~eV below the Gd L$_{2}$ edge. Above $T^\prime$, the (0~3.75~0) reflection is replaced by an incommensurate reflection (0~4-$\tau $~0) and in a small transition regime around $T^\prime$ coexistence of commensurate and incommensurate reflections is observed (Fig.~\ref{fig7}). A similar behavior was observed when measured on cooling from above $T_N$ (not shown), which demostrates that the observed (0~3.75~0) intensity above $T_{N}^{Gd}$ is not due to some ``frozen-in'' Gd-ordered volume fraction.

The temperature dependence of the (0~4-$\tau $~0) peak position results in a variation of $\tau = 0.16...0.245$ for the temperature range $T^\prime < T < T_{N}^{Mn}$ (Fig.~\ref{fig6}, top, inset), in close agreement with the previously reported XRD data\cite{Kimura:2004} for the Mn propagation vector $\tau ^{Mn}$. Apparently, this weak reflection can be used to track $\tau ^{Mn}$. Guided by the previous work on Tb- and DyMnO$_{3}$, we associate the weak reflections observed above $T_{N}^{Gd}$ with the Mn-induced ordering of the Gd moment with the same propagation vector, $\tau ^{Mn}$. This suggests a Mn-ordering propagation vector $\tau ^{Mn} = 1/4$ in the temperature range $T_{N}^{Gd} < T < T^\prime$.  

The observed temperature dependence of $\tau ^{Mn}$ is quite unusual, since the direction of the jump of $\tau ^{Mn}$ from 0.16 to 0.25 on cooling through $T^\prime$ is opposite to the trend observed above that temperature. Notably, lacking the present evidence for $\tau ^{Mn} = 1/4$ odering below $T^\prime$, a simple A-type stacking of Mn moments has been suggested previously for this $T$ region.\cite{Kimura:2003} Based on the present results we can not exclude a coexistence of A-type with $\tau ^{Mn} = 1/4$ stacking, where both would be related to different crystallographic components of the Mn magnetic moments. This open question, however, does not affect the following discussion.

In addition to the (0~3.75~0) we measured the temperature dependence of the (0~1/4~4) reflection for two different azimuths, namely with the $a$ axis parallel or perpendicular to the scattering plane. In the first case (scattering plane very close to \textit{ac}), the (0~1/4~4) intensity drops continuously on heating (Fig.~\ref{fig5}, solid symbols). Here, too, the reflection can be tracked up $T^\prime$ due to the Mn-induced polarization of Gd moments. No intensity jump is observed around $T$*. However, when changing the azimuth by 90$^\circ$ ($a$ perpendicular to the scattering plane, scattering plane very close to \textit{bc}), the temperature dependence of the (0~1/4~4) intensity shows a jump by 40{\%} upon heating through $T$* = 4.5 K (Fig.~\ref{fig6}). This azimuth dependence suggests that the intensity increase above $T$* is associated with an increase of the $b$ component of the Gd moment. For $a$ parallel to the scattering plane this signal is suppressed, since in this geometry $b$ is almost perpendicular to the scattering plane (suppressing any XRMS signal from the $b$ component in the $\sigma\rightarrow\pi^{\prime}$ channel). Assuming that the Gd moment is confined to the \textit{bc} plane, the observed intensity jump would correspond to an increase from $\mu_{b} \approx$ 0.45 to 0.66~$\mu_{c}$ at the $T$* transition (again considering the projection of the ordered moment onto the scattered beam ($\theta $ $\approx 25^\circ$)). From the 20\% jump height observed for the (0~3.75~0) reflection a somewhat larger increase to $\mu_{b} \approx 0.9~\mu_{c}$ is inferred. The origin of this discrepancy is unclear. The basic conclusions, however, are unaffected.

\subsubsection{In-field measurements}

The (0~1/4~4) reflection was further studied in an applied magnetic field perpendicular to the scattering plane. With the $a$ axis parallel to the scattering plane, the field direction was almost parallel to $b$ with a tilt of about   4.5~$^\circ$ towards $c$. With such a small deviation we expect a behavior practically identical to the ideal case $\mathbf{H}\| b$. 

\begin{figure}[bt!]
\begin{center}
\includegraphics[scale = 0.7]{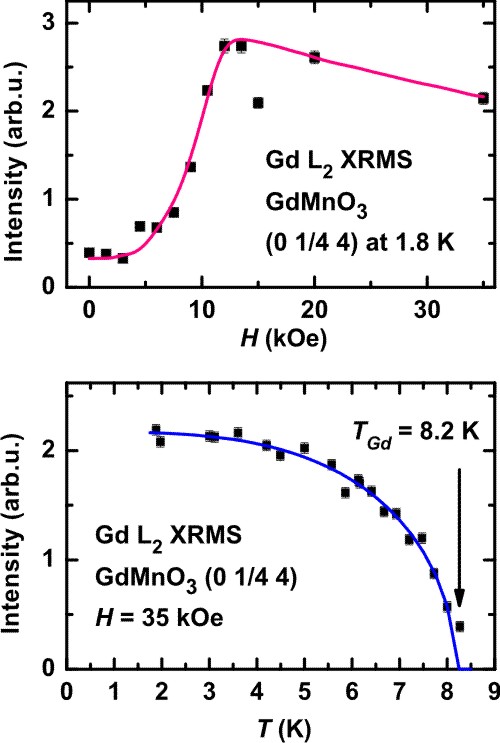}
\caption{(Color online)(top) Magnetic-field dependence at 1.8 K of the (0~1/4~4) Bragg reflection intensity and (bottom) temperature dependence in a 35~kOe applied field. Lines are guides to the eye. These data have been measured at the XmaS beamline (ESRF).}
\label{fig8}
\end{center}
\end{figure}

When a magnetic field is applied at $T $= 1.8~K, the (0~1/4~4) intensity exhibits a steep increase around 10~kOe (Fig.~\ref{fig8}, top). At 12~kOe it reaches a maximum of about 10 times the zero-field intensity, pointing to a significant field-induced modification of the Gd ordered structure. Higher fields lead to a minor decrease of the intensity. Obviously, this transition is related to a metamagnetic feature observed in magnetization measurements.\cite{Kimura:2005} 

We suppose that in the Gd zero-field ground state -- while modulated along $b$ with $\tau ^{Gd} = 1/4$ -- nearest neighbor Gd along $c$ are stacked antiferromagnetically, which would lead to the observed A- and F-type magnetic Bragg reflections. In applied magnetic fields $H > 10$~kOe along $b$ this ordering is modified, however, conserving the periodicity $\tau ^{Gd} = 1/4$. Probably, the stacking of nearest neighbor Gd along $c$ becomes ferromagnetic, which would lead to a strong increase of F-type magnetic reflections like (0~1/4~4). The temperature dependence of the (0~1/4~4) intensity in a 35~kOe applied field (Fig.~\ref{fig8}, bottom) indicates a vanishing Gd order above a transition temperature of $T_{N}^{Gd}$(35~kOe)~=~8.2~K.\cite{have:1}

We note that both the field-induced and temperature dependent transitions, observed for the Gd magnetic ordering, coincide with the phase boundaries between the ferro- and the paraelectric phases of GdMnO$_{3}$ reported before.\cite{Kimura:2005} Our data therefore suggest that a specific field-modified Gd moment arrangement is related to the occurrence of ferroelectricity in GdMnO$_{3}$. However, the comparison to the other ferroelectric members of the \textit{R}MnO$_{3}$ series and recent theoretical results\cite{Xiang:2008,Malashevich:2008} suggest that it is probably not the Gd ordering but the Mn that is responsible for the ferroelectricity in GdMnO$_{3}$. Previously, it has been argued that ferroelectricity in GdMnO$_{3}$ stems from a $\tau ^{Mn} = 1/4$ phase, analogous to the magnetic-field induced cycloidal $\tau ^{Mn~} = 1/4$ phase in TbMnO$_{3}$ that also exhibits a ferroelectric polarization 
$\mathbf{P}\| a$.\cite{Aliouane:2008,Arima:2005} In this picture, the field-induced ferroelectricity in GdMnO$_{3}$ would indicate a modification of the Mn magnetic ordering, in addition to the observed change of the Gd ordering, in applied field. Unfortunately, no detailed information on the Mn magnetic ordering in the ferroelectric phase is available. 

The previous results on Tb- and DyMnO$_{3}$ suggest a significant interplay between the rare earth and the Mn magnetism. Therefore, we suggest the following scenario for GdMnO$_{3}$. The incommensurate Mn ordering, $\tau ^{Mn} = 0.245${\ldots}0.16 observed below $T_{N}^{Mn}$ is a simple spin-density wave (SDW) that locks-in with a first-order transition at $T^\prime$ to give $\tau ^{Mn} = 1/4$. The latter phase is probably still a SDW, i.e., not cycloidal. On cooling through the Gd ordering transition, between $T_{N}^{Gd}$ and $T$* a cycloid on the Mn appears briefly, most probably with $\tau ^{Mn} = \tau ^{Gd} = 1/4$. However, it is unstable and disappears again below $T$*. When a magnetic field $H > 10$~kOe along $b$ is applied, another Gd moment stacking is realized. We speculate that this stacking might stabilize a cycloidal ordering of the Mn moments, leading to the observed ferroelectricity in GdMnO$_{3}$. In such a scenario, the Gd magnetic ordering would be very important for stabilizing the ferroelectric phase of GdMnO$_{3}$.

\section{Conclusions}

We have shown that the rare earth ions play an important role in determining the ferroelectric properties of        DyMnO$_{3}$ and (most probably also of) GdMnO$_{3}$. In DyMnO$_{3}$ we find a Mn-induced incommensurate magnetic ordering of Dy which leads to an enhancement of the ferroelectric polarization by a factor of up to three. This polarization is suppressed in zero magnetic field by the individual ordering of Dy moments with $\tau ^{Dy} = 1/2$, below $T_{N}^{Dy}$. Application of a magnetic field $\mathbf{H}\| a$ below $T_{N}^{Dy}$ destroys the commensurate phase. This leads to a re-emergence of the Mn-induced ordering and a concomitant strong enhancement of the ferroelectric polarization. In GdMnO$_{3}$, the magnetic-field induced ferroelectric phase for $\mathbf{H}\| b$ is associated with a specific ordering of the Gd moments with $\tau ^{Gd}= 1/4$. Presumably, this ordering stabilizes a cycloidal Mn ordering, which in turn produces an electric polarization analogous to the magnetic-field induced $\tau ^{Mn}$=~$1/4$ phase in TbMnO$_{3}$. Thus, it turns out that for a full understanding of ferroelectricity in the rare earth manganites an understanding of the rare earth magnetism is very important. 

\hspace{0.1cm}

\section{Acknowledgements}

A.U.B. Wolter has been supported by the DFG under Contract No. SU229/8-1. S. Landsgesell thanks the DFG for financial support under contract AR 613/1-1. Construction of the beamline MAGS has been funded by the BMBF via the HGF-Vernetzungsfonds under Contract Nos. 01SF0005 and 01SF0006. We thank J.~Voigt, E.~Schierle and E.~Weschke for inspiring discussions.


\begin{thebibliography}{27}
\bibitem{have:1} We have no data above 8.2K, since at this point of the experiment the sample fell off the sample holder. We take this as additional evidence that the sample actually entered the paraelectric, weak ferromagnetic phase, in which the large magnetic field would exert torque to the sample.
\end{thebibliography}
\end{document}